\documentclass[aps,prd,12point,twocolumn,nofootinbib,showpacs,superscriptaddress]{revtex4-2}
\def\theequation{\arabic{section}.\arabic{equation}}
\usepackage{psfrag}
\usepackage{subfigure}
\usepackage{color}
\usepackage{mathrsfs}
\usepackage{graphicx}
\usepackage{amssymb, bm}
\usepackage{amsmath, amsthm}
\usepackage{epstopdf}
\usepackage{hyperref}
\usepackage{enumerate}
\usepackage{longtable}
\usepackage{amsmath}
\usepackage[normalem]{ulem}

\newcommand{\be}{\begin{equation}}
\newcommand{\ee}{\end{equation}}

\definecolor{pinegreen}{rgb}{0.0, 0.47, 0.44}

\begin{document}
\def\theequation{\arabic{section}.\arabic{equation}}

\title{Critical solutions of nonminimally coupled scalar field theory and 
first-order thermodynamics of gravity}


\author{Valerio Faraoni}
\email[]{vfaraoni@ubishops.ca}
\affiliation{Department of Physics \& Astronomy, Bishop's University, 
2600 College Street, Sherbrooke, Qu\'ebec, Canada J1M~1Z7}

\author{Pierre-Antoine Graham}
\email[]{Pierre-Antoine.Graham@USherbrooke.ca}
\affiliation{Department of Physics, Universit\'e de Sherbrooke, 2500 
Boulevard de l'Universit\'e, Sherbrooke, Qu\'ebec, Canada J1K~2R1}

\author{Alexandre Leblanc}
\email[]{Alexandre.Leblanc3@USherbrooke.ca}
\affiliation{Department of Physics, Universit\'e de Sherbrooke, 2500 
Boulevard de l'Universit\'e, Sherbrooke, Qu\'ebec, Canada J1K~2R1}

\begin{abstract} 

Analytical solutions of nonminimally coupled scalar field cosmology 
corresponding to critical scalar field values constitute a potential 
challenge to the recent first-order thermodynamics of scalar-tensor 
gravity (a formalism picturing general relativity as the zero-temperature 
equilibrium state for modified gravity). The critical solutions are 
unstable with respect to homogeneous perturbations, hence unphysical.

\end{abstract}



\maketitle

\section{Introduction}
\label{sec:1}
\setcounter{equation}{0}

The idea that the Einstein equations may not be fundamental and that 
gravity could, after all, be an emergent phenomenon, has been contemplated 
for a long time \cite{Sakharov, Visser:2002ew, Volovik03, Barcelo:2005fc, 
Padmanabhan:2008wi, Padmanabhan:2009vy, Hu:2009jd, Verlinde:2010hp, 
Carlip:2012wa, Giusti:2019wdx}. Probably the most significant step in the 
emergent gravity program occurred when Jacobson derived the Einstein 
equations purely with thermodynamical considerations 
\cite{Jacobson:1995ab}.  A decade later, modified gravity was brought into 
this picture when the field equations of (quadratic) metric $f(R)$ 
gravity 
were derived with a similar thermodynamical procedure (here $R$ denotes 
the Ricci scalar of the spacetime metric $g_{ab}$ and $f(R)$ is a 
nonlinear function of $R$). A key idea was that general relativity (GR) 
constitutes a state of thermal equilibrium while modified gravity is an 
excited state \cite{Eling:2006aw}.  In this picture, the approach of 
modified gravity to GR would be a sort of relaxation to thermal 
equilibrium. The role of a ``viscosity of gravity'' in this process was 
suggested in \cite{Eling:2006aw} and clarified in later work 
\cite{Chirco:2010sw}.

In spite of many years of research on spacetime thermodynamics, 
little progress has really been made with respect to the original works 
\cite{Jacobson:1995ab,Eling:2006aw}. In particular, no equation describing 
the approach to the GR equilibrium state of modified gravity has been 
found and the order parameter (the ``temperature of gravity'') has not 
been identified. This state of ignorance about these key ingredients is 
very disappointing, especially in view of the fact that modified gravity 
is extremely popular nowadays and is the subject of intense research. The 
main motivation for studying alternative theories of gravity comes from 
cosmology: the standard $\Lambda$-Cold Dark Matter model of 
the universe based on Einstein theory invokes a completely {\em ad hoc} 
dark energy to explain the present acceleration of the universe discovered 
in 1998 with Type Ia supernovae (see \cite{AmendolaTsujikawabook} for a 
review). Many researchers dissatisfied with the idea of dark energy have 
resorted to modifying gravity at large scales instead 
\cite{Capozziello:2003tk,Carroll:2003wy}.  While this approach has its 
problems, it has also been shown as a proof of principle that it can 
explain the cosmic acceleration without introducing dark energy. The class 
of $f(R)$ theories of gravity is particularly popular for this purpose 
(see the reviews \cite{Sotiriou:2008rp, DeFelice:2010aj, Nojiri:2010wj}).

There are other independent and sound motivations to study deviations from 
GR. As soon as quantum corrections are introduced, gravity deviates from 
GR and exhibits higher order equations of motion or extra degrees of 
freedom. Likewise, the low-energy limit of the simplest string theory, the 
bosonic string, is an $\omega=-1$ Brans-Dicke theory 
\cite{Callan:1985ia, Fradkin:1985ys}  
(where 
$\omega$ is the parameter of the theory called ``Brans-Dicke coupling''  
\cite{Brans:1961sx}).  Returning to cosmology, but in the early (instead 
of late) 
universe realm, Starobinski inflation \cite{Starobinsky:1980te} seems to 
be the inflationary 
scenario currently favoured by observations \cite{Planck2} and is based on 
the 
Lagrangian density $R+\alpha \, R^2$ motivated by quantum 
corrections to the Einstein-Hilbert Lagrangian $R$.

From the point of view of emergent gravity it is intuitive that, as soon 
as extra degrees of freedom are added to the usual massless spin two modes 
of GR and are excited, one deviates from GR in what could be called a 
thermally excited state, and that GR represents a state of ``thermal 
equilibrium'' at lower ``temperature of gravity''. The problem is that 
this ``temperature of gravity'' and the ``approach to equilibrium'' are 
unknown.

A recent  alternative approach \cite{Faraoni:2021lfc,Faraoni:2021jri}  to 
the general picture 
of modified gravity as an excited state of GR  is completely 
different from Jacobson's thermodynamics of spacetime. It is minimalist in 
its assumptions and, contrary to spacetime thermodynamics, 
does not need to assume results from quantum field theory in curved 
spacetime (the Unruh temperature of uniformly accelerated observers) or 
horizon thermodynamics (the Bekenstein-Hawking entropy). The key idea is 
writing the field equations of modified gravity  in the form of 
effective Einstein equations with a right-hand side formed from all 
geometric terms different from the Einstein tensor (plus ``real'' matter, 
if present). This is possible when the gradient of the gravitational 
scalar field in the theory is timelike. It is a fact that, for the classes 
of theories examined 
(``first generation'' scalar-tensor gravity and viable Horndeski 
theories), this effective stress-energy tensor assumes the form of a {\em 
dissipative} fluid with a spacelike heat flux and shear and bulk 
viscosity \cite{Pimentel89,Faraoni:2018qdr,Quiros:2019gai, 
Giusti:2021sku}. The next step consists of applying to this dissipative 
fluid  Eckart's first-order 
thermodynamics \cite{Eckart40}. The latter is 
well-known to be non-causal and to suffer from instabilities 
\cite{Maartens:1996vi, Andersson:2006nr}  but nevertheless is still the 
most widely used model of 
dissipative fluid in relativity. We regard first-order thermodynamics of 
scalar-tensor gravity as a first step to be eventually 
replaced by a more realistic model. In spite of its crudeness, the 
first-order thermodynamics of scalar-tensor gravity has identified 
clearly a notion of ``temperature of gravity'' and has provided an 
equation describing the approach to the GR equilibrium state, or 
departures from it. Generic predictions of the formalism are 
\cite{Faraoni:2021lfc,Faraoni:2021jri}: a)~GR is the 
zero-temperature state of equilibrium;  b)~near spacetime 
singularities or near 
singularities of the effective  gravitational coupling, gravity is ``hot'' 
in the sense that it departs from GR, with ``temperature'' diverging at 
these singularities; c) 
the expansion of spacetime generally ``cools'' gravity bringing it closer 
to the GR equilibrium state; d)~theories in which the 
gravitational scalar field is non-dynamical ({\em e.g.}, cuscuton 
gravity \cite{Afshordi:2006ad,Afshordi:2007yx, 
Afshordi:2009tt,Bhattacharyya:2016mah, Iyonaga:2018vnu, Boruah:2017tvg, 
Boruah:2018pvq, Quintin:2019orx, Andrade:2018afh, Ito:2019fie, 
Ito:2019ztb, Gomes:2017tzd}) are also states of equilibrium 
at zero temperature \cite{Faraoni:2022doe}. This fact is explained by the 
fact that no extra degree of freedom with respect to GR is 
excited.\footnote{By 
contrast Nordstr\"om theory \cite{Nordstrom}, which contains only a 
dynamical scalar field 
(the conformal factor of the metric that is necessarily conformally flat 
in this theory), that is, much less freedom compared to GR, turns out to 
have negative temperature. This shows again that the ``temperature of 
gravity'' 
defined in \cite{Faraoni:2021lfc,Faraoni:2021jri} is relative to GR.}

The first-order thermodynamics of scalar-tensor gravity introduced in 
\cite{Faraoni:2021lfc,Faraoni:2021jri} for ``first 
generation'' scalar-tensor theories has been generalized to 
viable Horndeski gravity \cite{Giusti:2021sku} and then applied to 
scalar-tensor cosmology in \cite{Giardino:2022sdv}. Although the formalism is 
intriguing in many respects and explains certain features of scalar-tensor 
gravities, or of particular solutions of these theories, it is crucial to 
attempt to falsify the main ideas and predictions and look for places 
where the formalism could fail. In 
this regard, the inspection of particular theories \cite{Faraoni:2022doe}  
or of 
particular analytical solutions \cite{Faraoni:2022jyd, us}, could uncover 
corners where the formalism breaks down, eventually highlighting its 
limits of validity or leading to its rejection altogether. Certain 
critical solutions of nonminimally coupled  scalar 
field cosmology constitute a potential challenge to the first-order 
thermodynamics  because they are associated with 
an effective 
gravitational coupling $G_\mathrm{eff}$ that diverges identically through 
the entire cosmological dynamics and with an ill-defined effective 
temperature of gravity. Although it has been shown that the 
universe cannot pass through such singular points dynamically 
\cite{Starobinski80, Abramo:2002rn}, 
strictly speaking these special solutions evade this theoretical result 
because they are already at infinite $G_\mathrm{eff}$. We would like to 
study the stability of these solutions. If stable, one should worry about 
them because the effective temperature is undefined as  a result of the 
product of the divergent $G_\mathrm{eff}$ with a vanishing 
quantity, and the first order thermodynamical formalism could potentially 
break down.

In the next section we make these arguments explicit, introducing the 
field equations of nonminimally coupled scalar field cosmology and the key 
equations of the first-order thermodynamics of scalar-tensor gravity, as 
well as the critical solutions corresponding to infinite 
$G_\mathrm{eff}$. The stability of these solutions with 
respect to homogeneous perturbations is studied in Section~\ref{sec:3}, 
while Sec.~\ref{sec:4} contains our conclusions and a discussion. 

We adopt the notation of Ref.~\cite{Waldbook}, using units in which the 
speed of light $c$ is unity and the metric signature is ${-}{+}{+}{+}$. 
$\kappa \equiv 8\pi G$, $G$ is Newton's constant, and $V(\phi)$ is the 
potential of the gravitational scalar field $\phi$.

\section{Equations and unperturbed solutions}
\label{sec:2}
\setcounter{equation}{0}

The action for gravity with a nonminimally coupled scalar field $\phi$ is 
\be 
S= \int d^4 x \sqrt{-g} \, \left[ \left( 
\frac{1}{\kappa}-\xi \phi^2 \right) R - \frac{1}{2} \, \nabla^c\phi 
\nabla_c\phi - V(\phi) \right]  \,,\label{SNMC}
\ee
where $R$ is the Ricci scalar of the spacetime metric $g_{ab}$ with 
determinant $g$. The vacuum field equations are 
\begin{eqnarray}
G_{ab} &=& \frac{\kappa}{ 1-\kappa\, \xi \phi^2 } \Big[   
\nabla_a 
\phi \, \nabla_b \phi -\frac{1}{2} \, g_{ab} \nabla^c 
\phi \, \nabla_c \phi -\frac{V}{2}  \, g_{ab}  \nonumber\\
&&\nonumber\\
&\, &  +\xi \left( g_{ab} \Box -\nabla_a 
\nabla_b \right) \left( \phi^2 \right) \Big] \,,
\end{eqnarray}
\be 
\Box \phi -\frac{dV}{d\phi} - \xi R \phi =0 \,,
\ee
where $R_{ab}$ is the Ricci tensor, $G_{ab}= R_{ab}-g_{ab}R/2$ is the 
Einstein tensor,  and 
$\Box \equiv g^{ab} \nabla_a \nabla_b $ is the curved spacetime 
d'Alembertian. In 
the spatially flat Friedmann-Lema\^itre-Robertson-Walker (FLRW) geometry
\be
ds^2=-dt^2 +a^2(t) \left( dx^2 +dy^2 +dz^2 \right) \,,
\ee
the field equations assume the form
\begin{eqnarray}  
&& 6\left[ 1 -\xi \left( 1- 6\xi \right) \kappa \phi^2
\right] \left( \dot{H} +2H^2 \right) 
-\kappa \left( 6\xi -1 \right) \dot{\phi}^2     \nonumber\\
&&\nonumber\\
& & - 4 \kappa V + 6\kappa \xi \phi V' = 0 \, ,\label{fe1}
\end{eqnarray}

\begin{equation}  \label{fe2}
\frac{\kappa}{2}\,\dot{\phi}^2 + 6\xi\kappa H\phi\dot{\phi}
- 3H^2 \left( 1-\kappa \xi \phi^2 \right) + \kappa  V =0 \, ,
\end{equation}
\be  \label{fe3}
\ddot{\phi}+3H\dot{\phi}+\xi R \phi +V' =0 \,,
\ee
where an overdot denotes differentiation with respect to the 
comoving time $t$ and $H\equiv \dot{a}/a$ is the Hubble function. 

When $\phi \neq 0$, two of the three equations are independent and one can 
derive one of them from the other two. Since the cosmic scale factor 
$a(t)$ of a spatially flat universe only enters the equations of 
motion~(\ref{fe1})-(\ref{fe3}) in the combination $H \equiv \dot{a}/a$, 
the phase space consists of the three dimensions $\left( \phi, \dot{\phi}, 
H \right)$. However, Eq.~(\ref{fe2}) is a first order constraint 
(``Hamiltonian'' or ``scalar'' constraint \cite{Waldbook}) akin to an 
energy 
conservation equation in point particle mechanics, that 
forces the orbits of the solutions to move on a two-dimensional 
subset of this three-dimensional phase space. This feature is made evident 
by the fact that Eq.~(\ref{fe2}) yields $\dot{\phi}$ for any given 
pair of values of $\phi$ and $H$:
\begin{eqnarray}
\dot{\phi} \left( \phi, H \right) &=&  -6\xi H\phi \pm \Big[36 \xi^2 H^2 
\phi^2 +\frac{6H^2}{\kappa}\left(1-\kappa\xi \phi^2\right) \nonumber\\
&&\nonumber\\
&\, & -2V(\phi)  \Big]^{1/2} \,.\label{energysurface}
\end{eqnarray}
The double sign in front of the square root in Eq.~(\ref{energysurface}) 
describes the fact that the ``energy'' submanifold is usually comprised of 
two 
sheets which join at the points of the phase space where the argument of 
this square root vanishes (if the latter is negative there is a 
region of the 
phase space forbidden to the orbit of the solutions). This structure of 
the $\left( H, \phi, \dot{\phi} \right)$ phase space is discussed at 
length in \cite{Gunzig:2000ce} for nonminimally coupled scalar field 
cosmology and in \cite{Faraoni:2005vc} for more general scalar-tensor 
cosmology.

The nonminimal coupling of the scalar field introduces, in principle, the 
possibility of a negative effective gravitational coupling 
\cite{Linde:1979kf, Gurevichetal73, Pollock:1982tu, 
Hosotani:1986ga,Bamba:2014kza}
\be 
G_\mathrm{eff}=\frac{G}{1-\kappa \xi \phi^2} \,.\label{Geff}
\ee
Although there are rather compelling reasons to select conformal coupling 
$\xi=1/6$ \cite{Buchbinderbook, Buchbinder:1985ba, Buchbinder:1985ew, Markov86, 
Odintsov:1990mt, Muta:1991mw, Sonego:1993fw, Elizalde:1994im, 
Faraoni:1996rf}, we  discuss general, but positive, 
values of the nonminimal  coupling constant  $\xi$.  When $\xi>0$ there 
are two  critical values of the nonminimally coupled scalar  $\phi$,  
\be
\pm \phi_c \equiv \pm \frac{1}{ \sqrt{ \kappa \, \xi} } \,.
\ee
By defining 
\be
\psi \equiv 1-\kappa \xi \phi^2 \,,
\ee
the action~(\ref{SNMC}) is rewritten as the more familiar scalar-tensor  
action
\be 
S= \frac{1}{2\kappa} \int d^4 x \sqrt{-g} \, \left[ \psi R 
- \frac{\omega(\psi)}{\psi} \, \nabla^c\psi 
\nabla_c\psi - U(\psi) \right]  \,,\label{SNMC}
\ee
where $U(\psi)= V(\phi(\psi))$. 

The effective temperature ${\cal T}$ of 
scalar-tensor gravity in the context of Eckart's first-order 
thermodynamics for the 
effective fluid equivalent of the Brans-Dicke-like scalar $\phi$ was 
derived in \cite{Faraoni:2021lfc,Faraoni:2021jri}.  Its product with the 
effective thermal conductivity 
${\cal K}$ is \cite{Faraoni:2021lfc,Faraoni:2021jri}
\be
{\cal K}{\cal T}= \frac{ \sqrt{-\nabla^c\psi \nabla_c\psi}}{\kappa \psi} =
\frac{ 2\xi |\phi| \sqrt{ -\nabla^c\phi \nabla_c\phi}}{1-\kappa \xi\phi^2} 
\,. \label{temperature2}
\ee
This formula is derived in the general theory. Technically, in 
unperturbed FLRW cosmology, the spatial heat flux and shear viscosity are 
absent in order to preserve the spatial homogeneity and isotropy (but 
Eq.~(\ref{temperature2}) still makes sense, as discussed in 
\cite{Giardino:2022sdv}), but the viscous pressure and bulk viscosity 
remain and are given by \cite{Faraoni:2021lfc,Faraoni:2021jri, 
Giusti:2021sku,Giardino:2022sdv}
\be
P_\mathrm{viscous}=-\zeta \Theta \,, \quad\quad \zeta= -\frac{ {\cal 
K}{\cal T}}{3} 
\ee
according to Eckart's constitutive laws, where $\Theta = 3H $ is the 
expansion scalar and $\zeta $ is the effective bulk viscosity coefficient 
\cite{Faraoni:2021lfc,Faraoni:2021jri, Giusti:2021sku, Giardino:2022sdv}. 
As for ${\cal K}{\cal T}$, the critical solutions imply an 
ill-defined bulk viscosity coefficient $\zeta$. However, as soon as the 
critical FLRW universes are perturbed (as we do here), the effective 
temperature and bulk and shear viscosity coefficients are well-defined 
again.

Here we examine a few analytical solutions of FLRW cosmology sourced by a 
nonminimally coupled scalar field that describe  spatially 
flat FLRW universes with  constant Ricci scalar  
\be
R=6 \left( \dot{H}+2H^2 \right) \equiv 6C \label{R}
\ee
reported in \cite{Gunzig:2000ce, Faraoni:2001tq}. They are given by 
\be 
\phi = \pm \phi_c \equiv \frac{\pm 1}{\sqrt{\kappa \xi}}  
\label{phicritical}
\ee
and
\begin{eqnarray}
H_c(t) &=& \sqrt{ \frac{C}{2}} \, \tanh\left( \sqrt{2C}\, t \right)   
 \quad\quad  \mbox{for} \quad  C>0 \,,\label{Hcritical1}\\
&&\nonumber\\
H_c(t) &=& \frac{1}{2t}  \quad\quad  \mbox{for} \quad C=0 
\,,\label{Hcritical2}\\
&&\nonumber\\
H_c(t) &=&  - \sqrt{ \frac{|C|}{2}} \, \tan\left( \sqrt{2|C|}\, t \right)   
\quad\quad  \mbox{for} \quad C<0 \,,\nonumber\\
&& \label{Hcritical3}
\end{eqnarray}  
corresponding to the scale factors
\begin{eqnarray} 
a_c(t) &=& a_0 \sqrt{ \cosh \left( \sqrt{2C}\, t\right) }\,, \label{a1} \\
&&\nonumber\\
a_c(t) &=&  a_0 \, \sqrt{t} \,, \label{a2} \\
&&\nonumber\\
a_c(t) &=& a_0 \sqrt{ \cos \left( \sqrt{2C}\, t\right) } \label{a3}\,,
\end{eqnarray}
respectively, where $a_0$ is a constant. 

There are also critical de Sitter spaces\footnote{We refer to both  
exponentially expanding or contracting spatially flat FLRW universes (in 
comoving time) as ``de Sitter spaces'', although this terminology is 
normally restricted to expanding spaces.} with 
constant Hubble function \cite{Gunzig:2000ce}
\be
\phi=\pm \phi_c \,, \quad\quad H_c = \pm \sqrt{ \frac{C}{2}} \quad \quad 
\mbox{for} \quad C>0 \,, \label{Hcritical4}
\ee
or $a(t)=a_0 \, \mbox{e}^{ H_c\, t} $. 
de Sitter universes with constant scalar field are the only fixed points 
of the dynamical system~(\ref{fe1})-(\ref{fe3}) in phase space.  For 
$C=0$, these de Sitter universes degenerate into Minkowski spacetime.

All these solutions satisfy the field 
equations~(\ref{fe1})-(\ref{fe3}) 
provided that
\begin{eqnarray}
V_c & \equiv & V( \pm \phi_c) = 0 \,,\label{V_c}\\
&&\nonumber\\
V'_c & \equiv & \frac{dV}{d\phi} \Big|_{\pm \phi_c} = \mp 6\xi C \phi_c 
\,.
\end{eqnarray}


The critical values $\pm \phi_c$ of the scalar field are precisely those 
for which the effective gravitational coupling $G_\mathrm{eff}$ diverges. 
On the one hand, $\phi=$~const. implies ${\cal K}{\cal T}=0$; on the other 
hand,  
$\phi= \pm 
\phi_c$ implies an infinite $G_\mathrm{eff}= G \left[ 1- \left( 
\phi/\phi_c \right)^2 
\right]^{-1} $. As a result, the effective 
temperature~(\ref{temperature2}) of the 
effective dissipative fluid is ill-defined. This feature is  a puzzle for 
the first-order thermodynamics of scalar-tensor gravity. Its 
resolution comes from the realization that these analytical solutions 
of nonminimally coupled scalar field cosmology are 
unstable with respect to homogeneous perturbations. Therefore, they are 
not physically relevant since they will be destroyed by arbitrarily small 
perturbations and are not expected to occur in nature. 

The instability of anisotropic Bianchi universes  as $\phi \to \pm \phi_c 
$ 
was already established by Starobinski in \cite{Starobinski80} (and 
recovered in \cite{Abramo:2002rn}): $\phi$ can 
never pass dynamically through the critical values $\pm \phi_c$ where 
$G_\mathrm{eff}$ diverges, since the shear $\sigma_{ab}$ 
and the Kretschmann scalar $R_{abcd} R^{abcd} $ diverge there, for both 
classical and semiclassical $\phi$ \cite{Starobinski80}.   However,  for 
the exact solutions 
$\Big( \phi, H \Big)= \Big( \pm \phi_c , H_c(t) \Big)$ the 
situation is somehow different. The singularity $G_\mathrm{eff}=\infty$ is 
not approached dynamically but this  quantity is {\em identically}  
infinite. The 
effective temperature 
\be
{\cal K}{\cal T} = \frac{ 2\xi |\phi|  \sqrt{-\nabla^c\phi 
\nabla_c\phi}}{1-\kappa\xi \phi^2}  \sim 
\infty \cdot  0
\ee 
is ill-defined. Normally, a constant  $\phi$ reduces the theory to GR and 
${\cal K}{\cal T}$ to zero,  but singularities are ``hot'' in the sense 
that scalar-tensor gravity  deviates from GR in their proximity 
\cite{Faraoni:2021lfc, Faraoni:2021jri}.  
Moreover,  in the context of Eckart's effective 
thermodynamics, ``singularity'' 
should be intended either as a spacetime singularity or as a singularity 
of the effective gravitational coupling, as 
discussed in \cite{Faraoni:2021lfc, Faraoni:2021jri}.  
Starobinski's result suggests that the fine-tuned solutions $\phi=\pm 
\phi_c $, $ H=H_c(t)$ are dynamically unstable with respect to anisotropic 
perturbations; here we examine the stability of these critical solutions 
with respect to homogeneous perturbations.

\section{Perturbing the critical solutions}
\label{sec:3}
\setcounter{equation}{0}

Before discussing the stability of the critical 
solutions~(\ref{phicritical})-(\ref{Hcritical4}) of nonminimally coupled 
scalar field cosmology with respect to homogeneous perturbations, we note 
that established formalisms for more general inhomogeneous perturbations,  
such as the gauge-invariant formalism of 
Bardeen-Ellis-Bruni 
\cite{Bardeen:1980kt,Ellis:1989jt,Ellis:1989ju,Ellis:1990gi} in Hwang's 
version for modified gravity \cite{Hwang:1990am, 
Hwang:1990re, Hwang:1990jh, Hwang:1995bv, Hwang:1996bc, Hwang:1996xh, 
Hwang97,Noh:2001ia}, are not applicable where the effective 
gravitational coupling diverges. However, homogeneous perturbations make 
all the critical solutions unstable, which suffices to establish their 
instability.
 
The homogeneous perturbations are described  by
\begin{eqnarray}
\phi(t) &=& \pm \phi_c + \delta\phi(t) \,,\label{v1}\\
&&\nonumber\\
H(t) & = & H_c(t) +\delta H(t) \,. \label{v2}
\end{eqnarray}

It is difficult to make sense of a negative effective 
gravitational coupling \cite{Linde:1979kf, Gurevichetal73, Pollock:1982tu, 
Hosotani:1986ga,Bamba:2014kza}; moreover, the scalar 
field $\phi$ cannot cross 
dynamically the barriers $\phi = \pm \phi_c$ separating regions with 
positive effective gravitational coupling~(\ref{Geff}) from regions 
with negative $G_\mathrm{eff}$ \cite{Starobinski80, Abramo:2002rn}. 
Therefore, we impose that this coupling is always positive. This 
requirement means that 
perturbations $\delta\phi(t)$ of critical solutions with $\phi= +\phi_c$ 
must have $\delta\phi \leq 0$ to keep $ \phi \leq \phi_c$, while 
perturbations $\delta\phi(t)$ of critical solutions with $\phi=-\phi_c$ 
have $\delta \phi \geq 0$ to maintain $|\phi|\leq \phi_c $. As a 
consequence of this requirement, the effective 
temperature~(\ref{temperature2}) of the $\phi$-fluid is positive-definite. 
In short,
\be 
1-\kappa\xi \phi^2 
=1-\kappa\xi \left( \pm \phi_c+\delta\phi \right)^2 \simeq \mp 2\phi_c 
\delta\phi  \geq 0 \,. \label{eq:short}
\ee 

The perturbations in Eqs.~(\ref{v1}), (\ref{v2}) are substituted in 
Eqs.~(\ref{fe2})-(\ref{fe3}). Keeping only first order terms in 
$\delta\phi$ and $\delta H$, the field equations~(\ref{fe1})-(\ref{fe3})  
become
\begin{eqnarray}
&&\pm 6 \kappa \xi \phi_{c} H_{c} (\delta \dot{\phi} + H_c\delta 
\phi)+H_{c}(6 \kappa \xi \phi_{c}^{2}- 1) \delta H \nonumber\\
&&\nonumber\\
& &+ \kappa V^{\prime}_c \delta \phi=0 \,, \label{fep2_1}\\
&&\nonumber\\
&&\delta \ddot{\phi}+3 H_{c} \delta \dot{\phi}+\xi R \delta \phi \pm \xi 
\phi_{c} \delta R +V^{\prime \prime}_c \delta \phi=0  \,, \label{fep3_1} 
\end{eqnarray}
with $\delta R=6(\delta \dot{H}+4 H_{c} \delta H)$ from Eq.~(\ref{R}). 
Using~(\ref{v2}) and~(\ref{phicritical}), Eq.~(\ref{fep2_1})  
becomes 
\be
\delta \dot{\phi} + \left[H_c - \frac{C}{H_c}\right] \delta \phi =0 
 \label{fep2}
\ee
when $H_c\neq 0$ (The special case $H_c = 0$ is treated in sec.~\ref{special_de_sitter}). The solution of this homogeneous ODE is given by 
\be
\delta \phi(t) = \exp \left\{ -\int dt \left[ H_c(t) 
-\frac{C}{H_c(t)} \right] \right\} \,. \nonumber
\ee
The integral is computed using~(\ref{R}), obtaining 
\be
\delta \phi(t) = \delta \phi_{0} \, a_{c} H_{c} \label{sol_fep2}.
\ee
where now the sign of the (small) constant $\delta\phi_0$ must be such 
that the effective gravitational coupling~(\ref{Geff}) remains positive.

One can then use Eq.~(\ref{fep2}) to remove derivatives of $\delta \phi$ 
from the first two terms of~(\ref{fep3_1}):
\begin{eqnarray}
\delta \ddot{\phi}+3 H_{c} \delta  \dot{\phi} 
&=& \left[-H_{c}+\frac{C}{H_{c}}\right]^2 \delta \phi 
+\left[-\dot{H}_{c}-\frac{C\dot{H}_{c}}{H_{c}^{2}}\right]  \delta \phi 
\nonumber\\
&&\nonumber\\
&\, & +3 H_{c}\left[-H_{c}+\frac{C}{H_{c}}\right] \delta \phi 
=  2C \delta \phi 
\end{eqnarray}
and rewrite  Eq.~(\ref{fep3_1}) as 
\begin{align}
\delta \dot{H}+4 H_{c} \delta H =\mp \frac{1}{6\xi\phi_c} \left( 6\xi 
C+2C+V^{\prime \prime}_c \right) \delta \phi \equiv - D_{\pm} \delta 
\phi\label{fep3_2}\,,
\end{align}
where the the source term has been moved to the right-hand side. The 
integrating factor associated with this inhomogenous ODE is 
\begin{eqnarray}
 \exp \left\{4 \int dt\ H_{c} \right\} &=& \exp \left\{4 \int dt\ 
\left(\frac{\delta \dot{\phi}}{\delta 
\phi}-\frac{\dot{H}_{c}}{H_{c}}\right) \right\} \nonumber\\
&&\nonumber\\
&\sim &  \left(\frac{\delta \phi}{H_{c}}\right)^{4} \,,
\end{eqnarray}
where Eqs.~(\ref{fep2}) and~(\ref{R}) have been used to write 
\be
\delta \dot{\phi}=\left(-H_{c}+2 
H_{c}+\frac{\dot{H}_{c}}{H_{c}}\right) \delta 
\phi=\left(H_{c}+\frac{\dot{H}_{c}}{H_{c}}\right) \delta \phi  \nonumber
\ee
leading to 
\be
  H_{c}=\frac{\delta \dot{\phi}}{\delta \phi}-\frac{\dot{H}_{c}}{H_{c}} 
\nonumber
\ee
if $H_c$ does not vanish identically. The general solution  
computed using Eq.~(\ref{sol_fep2}) reads 
\be
\delta H = \delta \phi_{0} \left(-\frac{D_{\pm}}{5} a_{c}\right)+\delta 
H_0 \left(\frac{1}{a_{c}^{4}}\right) \,, \label{sol_fep3}
\ee
where $\delta H_0$ is a (small) integration constant for the homogeneous 
solution of Eq.~(\ref{fep3_1}) while the first term of~(\ref{sol_fep3}) is 
a particular solution of Eq.~(\ref{fep3_1}). The 
solutions~(\ref{sol_fep2}) and~(\ref{sol_fep3}) depend on only two free 
initial conditions $\delta \phi_0$ and $\delta H_0$, which is consistent 
with the presence of the phase space constraint~(\ref{energysurface}) and 
the resulting dimensionality of the phase space.

For completeness, we report the effective temperatures and bulk viscosity 
coefficients for the perturbed FLRW critical solutions. We 
have\footnote{This expression is only valid for attractive gravity and 
the absolute value making ${\cal K}{\cal T}$ positive-definite comes from 
the 
restriction~(\ref{eq:short}) on the sign of $\delta \phi$.}
\be
{\cal K}{\cal T} = \frac{2\xi |\phi|\sqrt{ 
-\nabla^c\phi\nabla_c\phi} }{1-\kappa\xi \phi^2  } = \frac{1}{\kappa} 
\Bigg| \frac{\delta\dot{\phi}}{ \delta\phi}\Bigg|  \label{gen_KT}
\ee
to first order, where in the last line we used the 
inequality~(\ref{eq:short}). Using Eq.~(\ref{fep2}), 
this yields
\be
{\cal K}{\cal T}= -3\zeta = \frac{1}{\kappa} \Big| \frac{C}{H_c}-H_c \Big| \,, \label{gen_KT2}
\ee
which is independent of the perturbation to this order, while the shear viscosity coefficient in the general 
theory is $\eta 
=3\zeta/2$ \cite{Faraoni:2021lfc,Faraoni:2021jri}.

We now study the stability of the specific critical 
solutions~(\ref{phicritical})-(\ref{Hcritical4}).

\subsection{Critical solution~I ($C=0$)}

\begin{figure}
\includegraphics[scale=0.5]{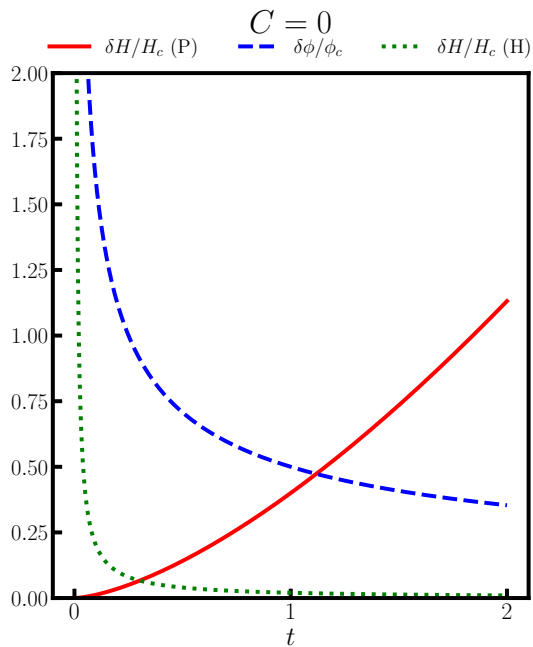}
\caption{\label{fig:1} Ratios of perturbation to unperturbed quantity for  
scalar field and Hubble function for the critical solution  with 
$C=0$. P and H denote, respectively, the particular solution of the 
inhomogeneous equation for $\delta H$ and the solution of the 
corresponding homogeneous equation. For illustration, the parameter values 
$a_0 = 1$, $\delta  \phi_0 = 1$, $D_\pm = -1$ and $\delta H_0 = 1/100 
$ have been used. To simplify the representation, we took 
$\phi_c = 1$ even when $\delta \phi > 0$.} 
\end{figure}

Inserting the critical solutions~(\ref{Hcritical2}) and (\ref{a2}) in 
Eqs.~(\ref{sol_fep2}) and (\ref{sol_fep3}) yields  
\begin{align}
&\delta H=\delta 
\phi_{0}\left(-\frac{D_\pm}{5}a_{0}\sqrt{t}\right)+\delta H_0 \left(\frac{1}{a_0^4 t^{2}}\right) \,,  \\ 
&\delta \phi  = \delta \phi_{0} \frac{a_{0}}{2 \sqrt{t}} \,.
\end{align}
For illustration, the ratios of these perturbations to the unperturbed 
critical quantities are  displayed in Fig.~(\ref{fig:1}) for specific 
parameter values, making it clear that at least one 
ratio always diverges at late times rendering the critical solutions for 
$C=0$ unstable. Here, the contribution of the particular solution 
of the inhomogeneous equation to the ratio 
$\delta H/H_c$ grows like $t^{3/2}$ and diverges as $t \to \infty$. 

${\cal K}{\cal T}$ and the bulk viscosity for this solution, 
determined with Eqs.~(\ref{gen_KT2}), read
\be
{\cal K}{\cal T} = -3 \zeta = \frac{1}{2\kappa \, t}.\nonumber
\ee
The perturbed universe expands forever and both temperature and bulk 
viscosity coefficient vanish in the late-time limit $t\to +\infty$ (they 
diverge near the Big Bang as $t\to 0^{+}$, but this limit is not 
meaningful since $\delta H/H_c \to \infty$ and the linear perturbation 
expansion breaks down).

\subsection{Critical solution~II ($C>0$)}

Combining Eqs.~(\ref{sol_fep2}) and (\ref{sol_fep3}) with 
the solutions~(\ref{Hcritical1}) and (\ref{a1}) yields

\begin{eqnarray}
\delta H &= & \delta \phi_{0}\left(-\frac{D_{\pm}}{5} a_{0} \cosh^{1 / 
 2}\left[\sqrt{2 C}t\right]\right)\nonumber\\
&&\nonumber\\
&\, &  +\delta H_0 
\left(\frac{1}{a_0^4}\text{sech}^{2}\left[\sqrt{2 C}t\right]\right) 
\,,\\
&&\nonumber\\
\delta \phi & = & \sqrt{\frac{C}{2}}  \delta \phi_{0} a_{0} \frac{\sinh 
\left[\sqrt{2 
C}t\right]}{\cosh^{1 / 2}\left[\sqrt{2 C} t\right]} \,.
\end{eqnarray}

\begin{figure}
\includegraphics[scale=0.5]{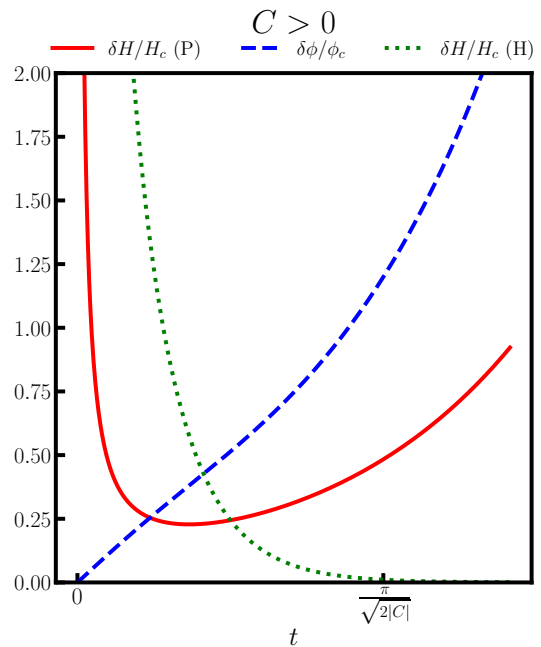}
\caption{\label{fig:2} Ratios of the perturbation to unperturbed quantity 
for   
scalar field and Hubble function for the critical solution with 
$C > 0$. P and H denote the particular solution of the 
inhomogeneous equation for $\delta H$ and the solution of the 
corresponding homogeneous equation. The parameter values 
$a_0 = 1$, $\delta  \phi_0 = 1/2$, $D_\pm = -1$, $\delta H_0 = 1 
$ and $C=1$ have been used. } 
\end{figure}

As shown in Fig.~\ref{fig:2}, both the particular part of $\delta H$ and 
$\delta  \phi$ lead to divergences for $t \to \infty$. Therefore, the 
solution for $C>0$ is unstable. 

The associated effective temperature and bulk viscosity are 
\be
{\cal K}{\cal T} = -3 \zeta = \frac{ \sqrt{ 2C }}{\kappa} \, 
\frac{
\left[ 1+\cosh^2 \left(\sqrt{2C}\, t\right) \right]}{\sinh\left( 
2\sqrt{2C} \, t \right)}.\nonumber
\ee
In the late-time limit $t\to+\infty$, 
\be
{\cal K}{\cal T} \to \frac{1}{\kappa} \, \sqrt{ \frac{C}{2}}. \nonumber
\ee
It seems that, ${\cal K}{\cal T}$ decreases asymptotically towards a 
constant at 
$t \to \infty$, which is consistent with the idea that expansion ($C>0$ 
corresponds to an expanding universe) ``cools'' gravity 
\cite{Faraoni:2021lfc, Faraoni:2021jri}, but ${\cal K} {\cal T}$ stops 
before reaching the GR state of equilibrium. All this is 
made irrelevant by the fact that the solution is unstable.  
The same value of ${\cal K}{\cal T}$ is obtained at all times for the 
critical de Sitter space in Sec.~\ref{special_de_sitter}.

\subsection{Critical solution~III ($C<0$)}

Substituting now the solutions~(\ref{Hcritical3})-(\ref{a3}) in 
Eqs.~(\ref{sol_fep2})-(\ref{sol_fep3}) leads to the  
perturbations
\begin{eqnarray}
\delta H & = & \delta \phi_{0}\left(-\frac{D_{\pm} a_{0}}{5}  \cos^{1 / 
2}\left[\sqrt{2 C}t\right]\right)\nonumber\\
&&\nonumber\\
&\, &  +\delta H_0 
\left(\frac{1}{a_0^4} \, \text{sec}^{2} \left[\sqrt{2 C}t\right]\right) 
\,,\\
&&\nonumber\\ 
\delta \phi & = & -\sqrt{\frac{C}{2} }\, \delta \phi_{0} a_{0}  \frac{\sin \left[\sqrt{2 
C}t\right]}{\cos^{1 / 2}\left[\sqrt{2 C} t\right]} \,.
\end{eqnarray}

\begin{figure}
\includegraphics[scale=0.5]{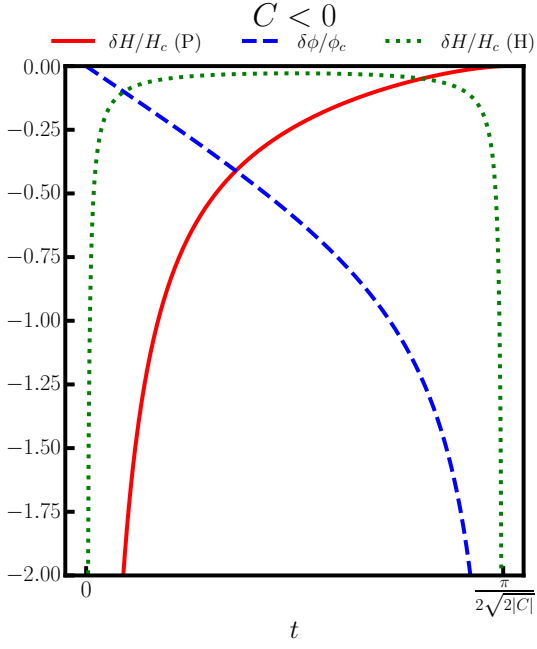}
\caption{\label{fig:3}  Same parameters as Fig.~\ref{fig:1} for the 
perturbations of  
the critical solution with $C = -1$.} 
\end{figure}

Figure~\ref{fig:3} shows how the homogenous part of the 
solution for $\delta H$ and $\delta \phi$ diverges for $t \to 
\pi/(2\sqrt{2C})$ and the solution is unstable. 

Again, $ {\cal K}{\cal T} $ and $\zeta$ are computed, giving 
\be
{\cal K}{\cal T} = -3\zeta = \frac{\sqrt{2|C|}}{\kappa}\,  \Bigg| 
\frac{3\cos^2\left( \sqrt{2|C|} \, t \right)-1}{\sin\left( 2\sqrt{2|C|} \, 
t \right) }\Bigg|\nonumber
\ee
and we have
\begin{eqnarray}
{\cal K}{\cal T} &\to &  + \infty \quad \mbox{as} \quad t\to 0 \,,\nonumber\\
&&\nonumber\\
{\cal K}{\cal T} &\to & +\infty \quad \mbox{as} \quad t\to \pm 
\frac{\pi}{2\sqrt{2|C|}} \,,\nonumber\\
&&\nonumber\\
{\cal K}{\cal T} &\to &  0 \,, \quad \mbox{at} \quad t = \pm 
\frac{\pi}{6\sqrt{2|C|}} \,.\nonumber
\end{eqnarray}

\subsection{Critical de Sitter spaces ($C>0$)}
\label{special_de_sitter}

\begin{figure}
\includegraphics[scale=0.5]{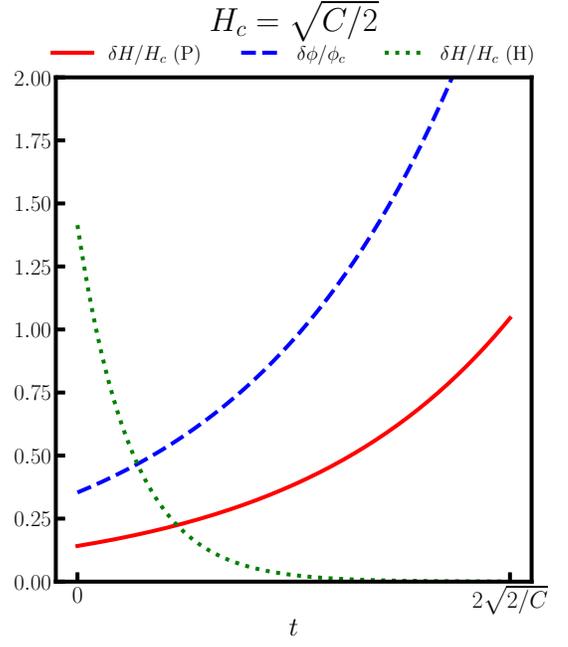}
 \caption{\label{fig:4} Same parameters as Fig.~\ref{fig:2} for the 
perturbations of  
the critical expanding de Sitter space.} 
\end{figure}

\begin{figure}
\includegraphics[scale=0.5]{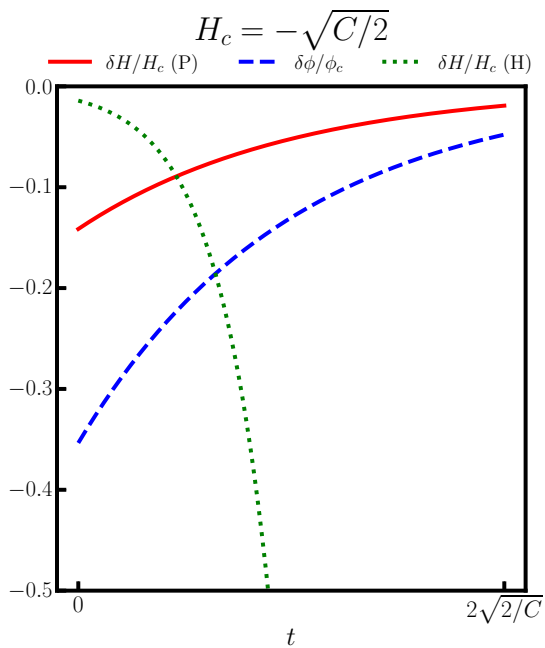}
\caption{\label{fig:5} Same parameters as Fig.~\ref{fig:2} for the 
perturbations of the critical contracting de Sitter space.} 
\end{figure}

For the special de Sitter spaces obtained for $C>0$ and given by 
Eq.~(\ref{Hcritical4}), the perturbations are 
\begin{eqnarray}
\delta H & = &\delta \phi_{0}\left[ -\frac{D_{\pm} a_0}{5} \, \exp 
\left(\pm\sqrt{ \frac{C}{2}} \, t\right)\right] \nonumber\\
&&\nonumber\\
  &\, & + \delta H_0 \left[ \frac{1}{a_{0}^{4}} \, \exp \left(\mp 4 
\sqrt{\frac{C}{2}} \, t \right)\right] \,,\\
&&\nonumber\\
\delta \phi &=& \pm \sqrt{\frac{C}{2}} \, \delta \phi_{0} \, a_{0} \exp 
\left(\pm \sqrt{\frac{C}{2}} \, t\right),
\end{eqnarray}
where the new $\pm$ sign represents the two possible signs for $H_c$ and 
is  independent of the sign appearing in $D_{\pm}$. These perturbations
are illustrated in Figs.~\ref{fig:4} and~\ref{fig:5}.  In each case, the 
exponential behavior is different for the homogeneous  and particular 
components of the solution for $\delta H$. Since there  is always one 
diverging exponential for $t \to \infty$, both solutions are 
unstable.

${\cal K}{\cal T}$ and the bulk viscosity coefficient are 
time-independent,  
\be
{\cal K}{\cal T} = -3 \zeta = \frac{1}{\kappa} \sqrt{\frac{C}{2}} \nonumber
\ee
These de Sitter spaces with constant effective 
temperature would be interpreted as metastable states, similarly 
to other special solutions discussed in \cite{Faraoni:2022jyd}.

If $C=0$, these de Sitter universes degenerate into a 
Minkowski spacetime for which Eq.~(\ref{fe1}) (with $V_c=0$, $V_c'=0$) 
gives  $\delta \dot{H}=0$, while Eq.~(\ref{fe3}) yields 
\be
\delta  \ddot{\phi} +V_c'' \, \delta \phi =0 \,.
\ee
The scalar field perturbation is unstable if $V_c'' \leq 0$ and would seem 
to be stable if $V_c''>0$ (that is, if the potential has a  
minimum with zero value at $\pm \phi_c$), in which case 
\be
\delta \phi(t) = A_0 \cos\left( \sqrt{V_c''} \, t \right) 
+ B_0 \sin\left( \sqrt{V_c''} \, t \right) \,.
\ee
However, the discussion at the 
beginning of this section informs us that, in order to keep the 
effective gravitational coupling non-negative, it must be $\delta \phi 
\leq 0$ 
for critical solutions with $\phi=\phi_c+\delta\phi$ and $\delta \phi \geq 
0$ for those with $\phi= -\phi_c+\delta\phi$. In both cases, the 
requirement $G_\mathrm{eff} \geq 0$ implies that $A_0=B_0=0$ and also 
this Minkowski space turns out to be unphysical.

\smallskip
\section{Conclusions}
\label{sec:4}
\setcounter{equation}{0}

The critical scalar field solutions correspond to ill-defined temperature 
and viscosity coefficients in the first-order thermodynamics of 
scalar-tensor gravity recently developed and one would like to 
understand their role. We have studied their stability with respect to 
homogeneous perturbations. Perturbed 
solutions of different type have different ${\cal K}{\cal T}$ close to the 
critical solutions. More precisely,  ${\cal K}{\cal T}$ is  
independent of $\delta \phi_0$ and $\delta H_0$, hence it is valid for 
arbitrarily small perturbations at any given time. While ${\cal K}{\cal 
T}$ is 
time-independent for the critical de Sitter solutions, it varies for 
the other solutions. This means that, approaching the $\phi = \pm 
\phi_c$ states from different directions in phase space, one obtains 
different values of ${\cal K}{\cal T}$, which is consistent with the fact 
that this quantity is undetermined at the 
critical scalar field value.

In order to keep the effective gravitational coupling $G_\mathrm{eff}$ 
positive near the critical field values $\pm \phi_c$, one must impose 
the condition  $\mp 2\phi_c 
\delta\phi  \geq 0 $ on the scalar field perturbations. The analysis of 
the previous section established 
that the critical solutions are unstable.  

Although we have reported effective temperature and bulk viscosity for the 
critical solutions (\ref{phicritical})-(\ref{Hcritical4}), their physical 
meaning is very questionable or irrelevant because  all these FLRW 
solutions are unstable 
and are destroyed already by homogeneous perturbations. Therefore, these 
critical solutions would not be realized in nature and they are of no real 
concern for the first order thermodynamics of scalar-tensor gravity. This 
remark is particularly important for de Sitter solutions with constant 
${\cal K}{\cal T}, \zeta$, and $\eta$  and for the late-time limit of the 
solution for $R>0$, which converges to a state where these quantities are 
also constant. If stable, these analytical solutions of nonminimally 
coupled scalar field cosmology would correspond to new states of 
equilibrium far away from GR, but they are unstable instead.

Having addressed this potential challenge, the first-order thermodynamical 
formalism will be developed further in future work.

\begin{acknowledgments}

This work is supported, in part, by the Natural Sciences \& Engineering 
Research Council of Canada (grant no.~2016-03803 to V.F.).

\end{acknowledgments}


\end{document}